\documentclass[aps,prl,preprint,floatfix,showpacs]{revtex4}
\usepackage{graphicx}
\begin{document}

\title{Voltage-controlled coded qubit based on electron spin}

\author{Pawel Hawrylak}
\affiliation{Quantum Theory Group, Institute for Microstructural
  Sciences, National Research Council of Canada, Ottawa, Ontario,
  Canada K1A OR6}

\author{Marek Korkusinski}
\affiliation{Quantum Theory Group, Institute for Microstructural
  Sciences, National Research Council of Canada, Ottawa, Ontario,
  Canada K1A OR6}

\begin{abstract}

We design and analyze a solid state qubit based on electron spin and
controlled by electrical means. The coded qubit is composed of a
three-electron complex in three tunable gated quantum dots. 
The two logical states of a qubit, $|0L\rangle$ and $|1L\rangle$, reside
in a degenerate subspace of total spin $S=1/2$ states. 
We demonstrate how applying voltages to specific gates changes the
one-electron properties of the structure, and show how electron-electron interaction
translates these changes into the 
manipulation of the two lowest energy states of 
the three-electron complex.

\end{abstract}


\maketitle

Considerable experimental and theoretical effort is currently
being applied towards the realisation of scalable quantum bits and
gates.
\cite{feynman,bennett,nielsen,vandersypen,bayer}
Electron spin-based qubits are considered to be viable candidates due
to demonstrated long coherence times \cite{awschalom,fujisawa,hanson}.  
For single quantum dots \cite{hanson,tarucha,ciorgaprb} it is now
possible to isolate, control, and exploit the spin properties of
electron at the single-spin level \cite{ciorgaprb,pioro}. 
Simultaneously, progress has been made in coupled quantum dot systems,
where control of electron numbers in individual dots and coupling
between the dots necessary to realise the building blocks for quantum
information processing
\cite{pioro,livermore,oosterkamp,duncan,holleitner,craig} 
 has been demonstrated. 
Despite this progress many obstacles need to be overcome before
elementary quantum operations on electron spin-based qubits\cite{brum,loss,ciorga,kane}  can be 
demonstrated. 
One of the fundamental challenges is the task of isolating a
single-electron spin and contacting it to perform single-qubit
operations. 
This challenge stems from the need to produce a localised and
switchable magnetic field required for spin manipulation. 
A number of solutions have been proposed \cite{kato,rashba,xiao,elzerman} to replace
the local magnetic field by locally engineered coupling ($g$-factor) 
to the constant magnetic field. 
The $g$-factor engineering has to be combined with equally challenging
spatial manipulation of electron position using surface gates. 
To date, even single-qubit operation of qubits based on electron spin
still awaits demonstration. 
An alternative solution to the problem of qubit operations on a single
spin was proposed by DiVincenzo, Lidar and collaborators \cite{divincenzo,mizel,bacon},
who suggested quantum computation with exchange interaction. 
The basic idea is to replace the two-level system based on the
single-electron spin with selected levels of a composite object
consisting of several spins. 
The manipulation of selected quantum levels proceeds not through
operations on single spins but through manipulation of the coupling
$J$ between neighbouring spins due to exchange interaction. 
For example, two qubit states $|0\rangle$ and $|1\rangle$ for a single
total spin $S=1/2$ can be identified with two opposite spin directions
($S_z$) up and down: $|0\rangle=|\downarrow\rangle$  
and $|1\rangle = |\uparrow\rangle$ . 
By contrast, in a coded qubit consisting of three localised electron
spins ($a$, $b$, $c$) we start by selecting a total spin $S_z$   
of the three spins, e.g., $|\downarrow\rangle$. 
A state with this direction can be realised by three different states:
$|\downarrow a, \downarrow b, \uparrow c\rangle$,  
$|\downarrow a, \uparrow b, \downarrow c\rangle$, and
$|\uparrow a, \downarrow b, \downarrow c\rangle$,
which differ by the position of the spin pointing up.  
We can group these three states into a single state with total spin 
$S=3/2$ and two orthogonal states with total spin $S=1/2$. 
Examples of the two orthogonal spin states forming the two logical
qubit states $|0L\rangle$ and $|1L\rangle$ are
$|0L\rangle = {1\over\sqrt{2}}(|\downarrow a, \downarrow b, \uparrow
c\rangle - |\downarrow a, \uparrow b, \downarrow c\rangle)$
and
$|1L\rangle = {1\over\sqrt{6}}(|\downarrow a, \downarrow b, \uparrow
c\rangle + |\downarrow a, \uparrow b, \downarrow c\rangle
-2 |\uparrow a, \downarrow b, \downarrow c\rangle$).

The properties of a three-electron complex in a single quantum dot,
including the two logical spin states discussed above, have been analyzed
previously\cite{hawrylak,ashoori}. 
The three-electron states are a product of the orbital and spin part. 
The spin part has been discussed in terms of the logical qubit
states. 
But it is precisely the orbital part, and specifically the one electron 
wavefunction, that can be manipulated using 
voltages, and hence it plays a crucial role in a coded qubit. 
In what follows we employ the methodology which allowed for a
quantitative understanding of a single dot with controlled number of
electrons \cite{kyriakidis} to study the energy levels and spin and
orbital wave functions of a realistic coded qubit. 
We show that there exist two low-energy states which map very well
onto the two logical qubit states, and that they can be manipulated by
applied voltages.  

The proposed device is shown in Fig.~\ref{fig1}a. 
It consists of a metallic gate defining three coupled lateral quantum
dots.  
The dots are defined by locally depleting the two-dimensional electron
gas (2DEG) at a distance $D$ below the surface. 
Application of a negative voltage creates a broad region depleted of 
electrons, with three identical minima. 
An additional gate (the control gate $V_x$), shown in red, is
positioned in-between the two adjacent dots 1 and 2.  
The gate $V_x$ lowers or increases the tunnelling barrier between the
two dots. 
As we will show, the effect of the gate is identical to the effect of
the $\sigma_x$ operation acting on logical qubit states. 
The $\sigma_x$ acting on logical qubit $|0L\rangle$ transforms it into
qubit $|1L\rangle$ and vice versa. 
The manifestation of the $\sigma_x$ operation is the energy splitting
$\Delta_x$ of the two degenerate logical qubit states.  
In order to have complete logical qubit rotation we need a second
gate. 
This gate, labelled $V_z$, plays the role of $\sigma_z$ spin
operation, and in our model it tunes the potential minimum of the
third dot. 
Finally, we need to precisely control the number of electrons $N=3$ in
our device. 
This can be accomplished by following the detailed design which
allowed to control the number of electrons in single and coupled
quantum dots \cite{ciorgaprb,pioro}.  

The operation of gates translates into the changes of a potential seen
by each of the three electrons in a coded qubit. 
The electrostatic potential $V_{E}(\vec{r})$ in the electronic 
plane due to the gates on the surface is given by an integral over the
potential $V_G(\vec{r})$ applied to the gates as
\cite{kyriakidis,davies}
\begin{equation}
V_E(\vec{r}) = \int{ {d\vec{r}'\over 2\pi}
|D|{V(\vec{r}')  \over (D^2+(\vec{r}-\vec{r}')^2)^{3/2}} },
\end{equation}
where $D$ is the distance between the surface and the 2DEG layer, and
$V_G(\vec{r})$ is the electrical potential on the surface,
corresponding to the appropriate voltage on the gates, and equal to
zero in the openings (the holes). 
In our calculation, we express all energies and distances in the units
of effective Rydberg, ${\cal R} = m^*e^4/2\varepsilon^2\hbar^2$, 
and effective Bohr radius, $a_B= \varepsilon\hbar^2/ m^*e^2$,
respectively.  
Here, $e$ and $m^*$ are the electronic charge and effective mass, 
respectively, $\varepsilon$ is the dielectric constant of the
semiconductor, and $\hbar$ is the Dirac's constant. 
For GaAs these units are $1 {\cal R}= 5.93$ meV, and $1 a_B = 97.9$
\AA $\sim$ 10 nm. 
The candidate for a coded qubit investigated here has the length of
the side of  the rectangular gate of $22.4$ $a_B$ (for better
visibility of the potential minima, in Fig.~\ref{fig1} we only show a
central part of the gate, with side length of 14 $a_B$). 
The diameter of the opening (the hole) in the gate is taken to be
$4.2$ $a_B$, the distance between the centers of each pair of holes is
$4.85$ $a_B$, and the distance between the gate and the 2DEG layers is
14 $a_B$. 
The voltage applied to the main gate corresponds to the electronic
potential energy $-eV=10$ ${\cal R}$ in the plane of the gates. 
As shown in Fig.~\ref{fig1}, the pattern of the holes in the surface
gate translates into three potential minima on the electronic plane. 
As shown in Fig.~\ref{fig1}a, if the voltage of the control gate
$V_x$ is set to zero, the three potential minima are identical. 
But if we apply voltage $-eV=-10$ ${\cal R}$, i.e., opposite to 
that of the main gate, the barrier between two of the dots is lowered
and the dots 1 and 2 are strongly coupled. 
Note that the two other tunnelling barriers are weakly affected by 
the control gate. 
The voltage applied to the gates couples to the charge of the electron
and determines the potential $V_E$ acting on each individual
electron. 
The potential determines the single-electron energies $E_n$ and wave
functions $\phi_n(x,y)$. 
The external voltages applied to control gates affect directly only
single-electron properties, and through the modification of
single-electron energies and wave functions - the states of  
the three-electron complex. 
To calculate the single-electron spectrum we discretize the area under
the gates, and define the single-electron states $\phi_n(x,y)$ on a
lattice $(x_i,y_j)$ with spacing $h$. 
Electrons move on a lattice, and their spectrum is described by the
tight-binding Hamiltonian:
\begin{equation}
(E_{i,j} + V_E(i,j))\phi_n(i,j) + 
\sum_{k,l} t_{i,j;k,l} \phi_n(k,l) = E_n\phi_n(i,j),
\label{eq1}
\end{equation}
where the site energy and hopping matrix elements are given by
$E_{i,j} = 4h^{-2}$ and $t_{i,j;k,l} =
-h^{-2}\delta_{k,i\pm1}\delta_{l,j\pm1}$, respectively. 
As a result we obtain a large, but sparse Hamiltonian matrix, which is
diagonalized using the conjugate gradient method. 
In Fig.~\ref{fig2}a we show the calculated energies $E_n$
corresponding to nine lowest single-particle states as a function of
the potential $V_x$ of the control gate. 
At zero bias the spectrum consists of three energy levels, the ground
state and two degenerate excited states, separated by a gap from the
rest of the spectrum. 
The corresponding three lowest single-particle wave functions at zero
bias are shown in Fig.~\ref{fig2}b. 
The states and the energy spectrum can be understood by considering a
linear combination of wave functions $f_m(x,y)$ localized on $m$-th
dot. 
The corresponding ground state can be written as 
$|0\rangle = {1\over\sqrt{3}}(f_1(x,y)+f_2(x,y)+f_3(x,y))$,
and the two degenerate excited states as 
$|1\rangle = {1\over\sqrt{6}}(f_1(x,y)+f_2(x,y)-2f_3(x,y))$ and
$|2\rangle = {1\over\sqrt{2}}(f_1(x,y)-f_2(x,y))$.
In all of these states the electron is delocalised and shared between
all dots. 
For negative gate voltages, i.e., when the barrier between the two
dots is lowered, the two excited states mix and energy levels split. 
When the voltage is zero, the three dots are identical, and the two
excited states are degenerate. 
As the barrier between the two dots is increased even further
(positive gate voltages), the energies of degenerate states are split
again. 
In this case, however, the splitting is not large. 
We attribute it to the fact that, in this regime, all interdot
barriers are already high and the three dots are almost isolated.  
Therefore, a further increase of the control gate voltage increases
the total energy of the system, but does not lead to a significant
symmetry breaking. 
Now that we understand the effect of gate voltages on a single
electron, we proceed to consider three electrons localized in our
three-dot potential. 
We describe simultaneously the spin and orbital three-electron states
in the language of second quantization as 
$|i\sigma,j\sigma',k\sigma''\rangle = 
c_{i\sigma}^+c_{j\sigma'}^+c_{k\sigma''}^+|0\rangle$. 
The operator $c_{i\sigma}^+$ ($c_{i\sigma}$) creates (annihilates) an
electron with spin $\sigma$ on the single-particle state $\phi_i$
calculated from Eq.~\ref{eq1}. 
The electron-electron interactions mix different three-electron
configurations for a given set of applied voltages. 
The mixing is governed by the matrix elements of the Hamiltonian,
which takes the form:
\begin{equation}
H = \sum_{i\sigma}E_{i\sigma}c_{i\sigma}^+c_{i\sigma}
+ {1\over2}\sum_{ijkl\sigma\sigma'}\langle i,j|V|k,l\rangle
c_{i\sigma}^+c_{j\sigma'}^+c_{k\sigma'}c_{l\sigma},
\label{mbham}
\end{equation}
where the energies $E_{i\sigma}(V_x,V_z)$ and matrix elements
$\langle i,j|V|k,l\rangle(V_x,V_z)$  of the Coulomb potential are
implicit functions of the applied voltages $V_x$, $V_z$. 
These matrix elements are independent of spin. 
They are calculated in real space as 
\begin{equation}
\langle i,j|V|k,l\rangle =
2h^3 \sum_{sp,uv}{
\phi_i(s,p)\phi_j(u,v)\phi_k(u,v)\phi_l(s,p) \over
\left[ (s-u)^2 + (p-v)^2 + d^2 \right]^{1/2}},
\end{equation}
with parameter $d$ accounting for the finite thickness of the electron
layer (in the following example we take $d = 0.2$ $a_B$).
To capture spin effects we generate all possible configurations of
three electrons on $N_S$ single-particle states and classify them by
total spin \cite{wensauer}. 
This allows us to construct the Hamiltonian matrix separately in the
spin $3/2$ and spin $1/2$ basis and diagonalize these matrices
numerically. 
Figure~\ref{fig3} shows the low-energy segment of the three-electron 
energy spectrum as a function of the voltage on the control gate $V_x$
(the energies are measured from the ground-state energy). 
The low-energy spectrum consists of two low-energy states
corresponding to the two states of the $S=1/2$ Hilbert space (black
lines), while the energy of the higher state corresponds to the
high-spin $S=3/2$ state (red line).  The fact that the two $S=1/2$ logical qubit 
states are the lowest energy states is  important as it facilitates the intialization
of a qubit. It is also a rather counterintuitive result as one might expect exchange to
favor spin polarized state. A detailed analysis of the correlations in
a coded qubit and analogies with Hubbard model will be presented elsewhere. 
The central result of this paper rests on the identification of
the two lowest energy $S=1/2$ states of a three electron complex
as the two logical states of the coded qubit. The two logical states 
  can be manipulated by applying gate
voltage $V_x$, which shifts the energy of the two levels, as seen in Figure~\ref{fig3}.
The voltage $V_x$ acts analogously to the $\sigma_x$ operation, while
the $\sigma_z$ operation can be implemented by applying voltage
$V_z$. 

The remaining concern is that the
two logical qubit states be well isolated from other states with the same total spin.   
The inset in Fig.~\ref{fig3}  shows the energy spectrum
over larger energy scale, with a significant gap in the energy spectrum between
the two logical qubit states and the remaining $S=1/2$ states.

In summary, we show that realistic calculations of a lateral gated
quantum dot device produce a voltage-tuneable qubit based on electron
spin.  The device should be able to combine long coherence time inherent to spin with
ease of operation inherent to electrical means. 
While the understanding of real-time operation of the qubit, its coupling to
environment, and design of complex qubit circuits remains to be investiagted, 
it appears that coded qubits based on electron spin are promising devices for
quantum information processing.

\section{Acknowledgments}
M.K. acknowledges financial support from the Natural
Sciences and Engineering Research Council of Canada and
P.H. acknowledges the support from the Canadian Institute for Advanced
Research. The authors thank A. Sachrajda and G. Austing for
discussions.

\newpage

\begin{figure}[h]
\caption{Cross-sectional view of the coded qubit realized on three
  coupled gated lateral quantum dots. 
  The grey rectangular gate contains three circular openings, which
  translate into minima of the electrostatic potential at the level of
  the two-dimensional electron gas. 
  The green gate can be used to shift the potential minimum of the dot
  underneath with respect to the two other dots. 
  The red gate is used to tune the tunnelling barrier between two of
  the dots. 
  If this gate is not polarized (a), the three dots are identical; if
  its potential is equal in value and opposite in sign to that of the
  main gate (b), the tunnelling barrier between the two dots below is
  lowered. }
\label{fig1}
\end{figure}

\begin{figure}[h]
\caption{
  (a). Energies of nine lowest single-electron levels of the three-dot
  system as a function of voltage applied to gate $V_x$. 
  (b) Wave functions corresponding to the three lowest single-electron
  energy levels of the three-dot system.}
\label{fig2}
\end{figure}

\begin{figure}[h]
\caption{ 
  Energies of the three lowest states of the three-electron coded qubit
  as a function of the voltage applied to the control gate $V_x$
  measured from the ground state. 
  Black lines show energies of total-spin-$1/2$ states identified as
logical qubit states, the red line  
  shows the energy of the spin-$3/2$ state. 
  Inset shows the gap between the two logical qubit states and the rest
of the spectrum as a function of the gate
  voltage.}
\label{fig3}
\end{figure}


\begin{thebibliography}{99}
\bibitem{feynman} 
  Feynman, R. P., Foundations of Physics {\bf 16,} 507 (1986).
\bibitem{bennett}  
  Bennett, C. H. and DiVincenzo, D. P., Nature {\bf 404,} 247 (2000).
\bibitem{nielsen}
  Nielsen, M. A., Knill, E., and Laflamme, R., Nature {\bf 396,} 52
  (1998). 
\bibitem{vandersypen} 
  Vandersypen, L. M., {\em et al.,} Nature {\bf 414,} 883 (2001). 
\bibitem{bayer}
  Bayer, M., {\em et al.,} Science {\bf 291,} 451 (2001).
\bibitem{awschalom} 
  Awschalom, D. D. and Kikkawa, J. M., Phys. Today {\bf 6,} 33 (1999).

\bibitem{fujisawa}
  Fujisawa, T., {\em et al.,} Nature {\bf 419,} 278 (2002).

\bibitem{hanson}
  Hanson, R., {\em et al.,} Phys. Rev. Lett. {\bf 91,} 196802 (2003).

\bibitem{tarucha}
  Tarucha, S., {\em et al.,} Phys. Rev. Lett. {\bf 77,} 3613 (1996).
\bibitem{ciorgaprb}
  Ciorga, M., {\em et al.,} Phys. Rev. B {\bf 61,} R16315 (2000).
\bibitem{pioro}
  Pioro-Ladriere, M., {\em et al.,} Phys. Rev. Lett. {\bf 91,} 026803
  (2003).
\bibitem{livermore}
  Livermore, C., {\em et al.,} Science {\bf 274,} 1332 (1996).
\bibitem{oosterkamp}
  Oosterkamp, T. H., {\em et al.,} Nature {\bf 395,} 873 (1998).
\bibitem{duncan}
  Duncan, D. S., {\em et al.,} Phys. Rev. B {\bf 63,} 045311 (2001).
\bibitem{holleitner}
  Holleitner, A. W., {\em et al.,} Science {\bf 297,} 70 (2001).
\bibitem{craig} 
  Craig, N. J., {\em et al.,} Science {\bf 304,} 565 (2004).

\bibitem{brum}
  Brum, J. A. and Hawrylak, P., Superlatt. Microstruct. {\bf 22,} 431
  (1997). 
\bibitem{loss}
  Loss, D. and DiVincenzo, D. P., Phys. Rev. A {\bf 57,} 120 (1998).
\bibitem{ciorga}
  Ciorga, M., {\em et al.,} Physica E {\bf 11,} 35 (2001).
\bibitem{kane}
  Kane, B. E., Nature {\bf 393,} 133 (1998).
\bibitem{kato}
  Kato, Y., {\em et al.,} Nature {\bf 427,} 50 (2004).
\bibitem{rashba}
  Rashba, E. I. and Efros, A. L., Phys. Rev. Lett. {\bf 91,} 126405
  (2003).
\bibitem{xiao}
  M. Xiao, I. Martin, E. Yablonovitch, and H. W. Jiang, 
  Nature {\bf 430,} 435 (2004).
\bibitem{elzerman}
  Elzerman, J. M., Hanson, R., Willems van Beveren, L. H., Witkamp,
  B., Vandersypen, L. M. K., and Kouwenhoven, L. P.,
  Nature {\bf 430,} 431 (2004).
\bibitem{divincenzo}
  DiVincenzo, D. P, {\em et al.,} Nature {\bf 408,} 339 (2000).
\bibitem{bacon}
  Bacon, D., {\em et al.,} Phys. Rev. Lett. {\bf 85,} 1758 (2000).
\bibitem{mizel}
  Mizel, A. and Lidar, D. A., Phys. Rev. Lett. {\bf 92,} 077903
  (2004).

\bibitem{hawrylak} 
  Hawrylak, P., Phys. Rev. Lett. {\bf 71,} 3347 (1993).
\bibitem{ashoori}
  Ashoori, R. C., Nature {\bf 379,} 413 (1996).
\bibitem{kyriakidis}
  Kyriakidis, J. {\em et al.,} Phys. Rev. B {\bf 66,} 35320 (2002).
\bibitem{davies}
  Davies, J. H., Larkin, I. A., and Sukhorukov, E. V., 
  J. Appl. Phys. {\bf 77,} 4504 (1995).
\bibitem{wensauer}
  Wensauer, A., Korkusinski, M., and Hawrylak, P., 
  Solid State Commun. {\bf 130,} 115 (2004).

\end{thebibliography}
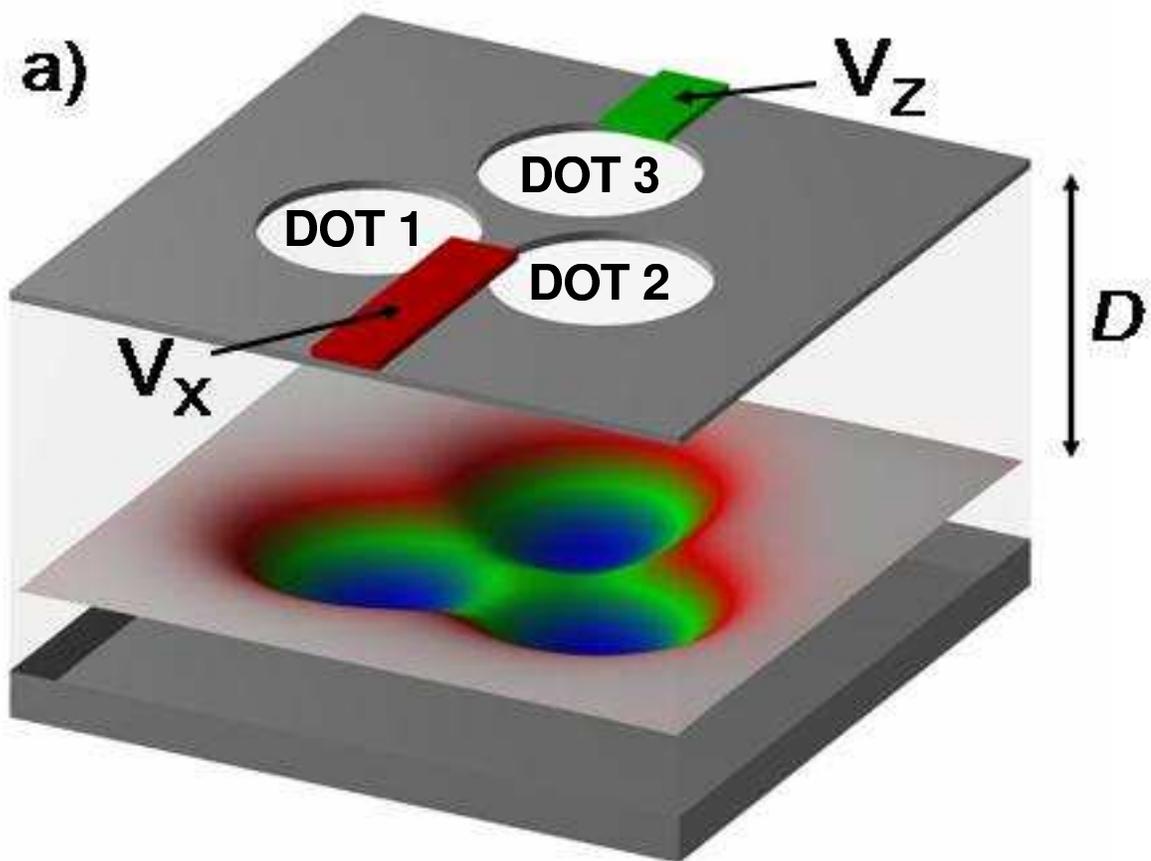
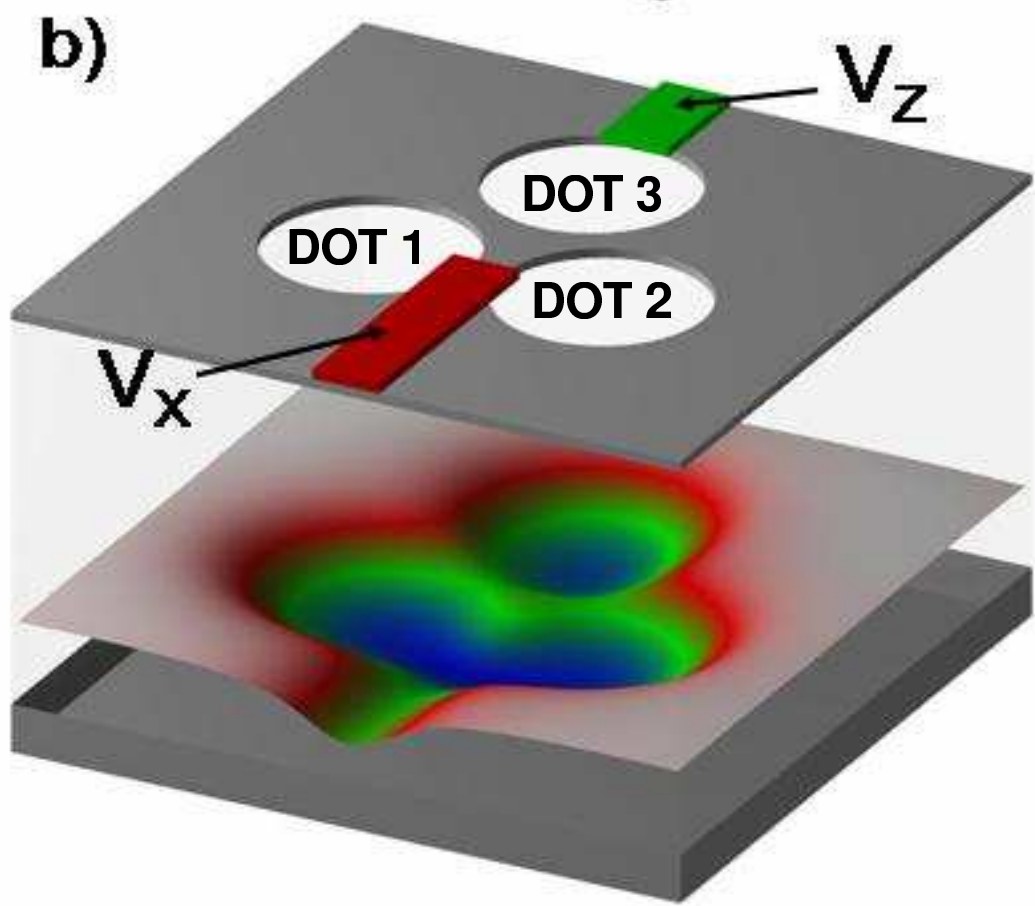
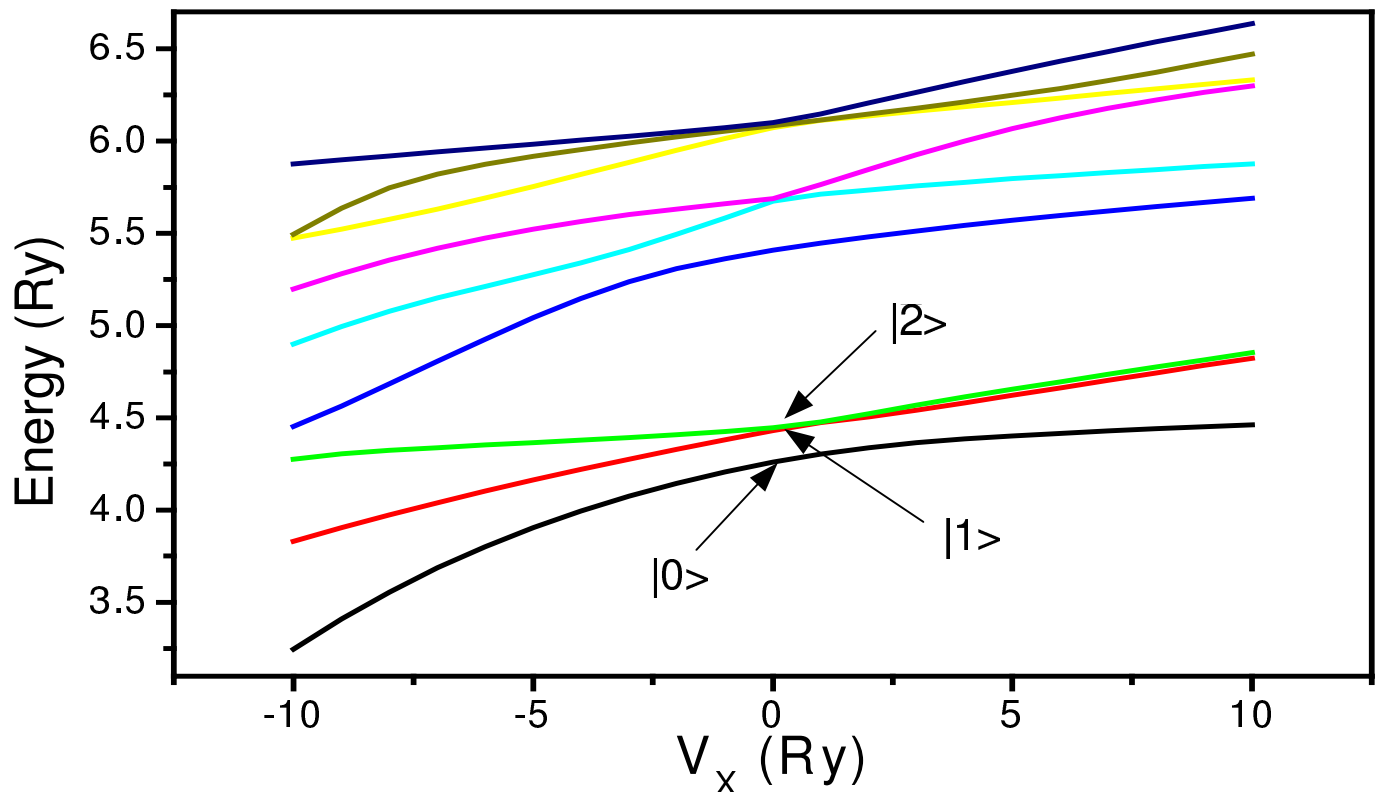
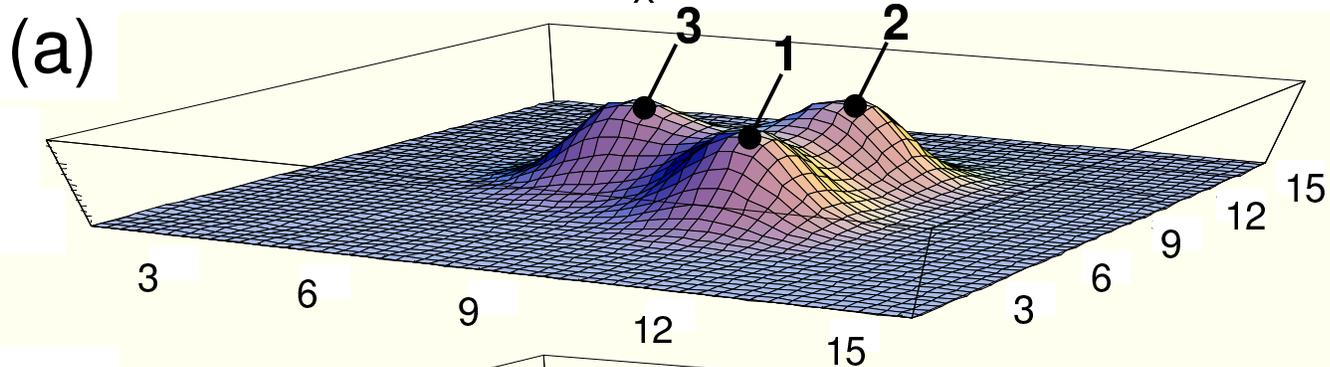
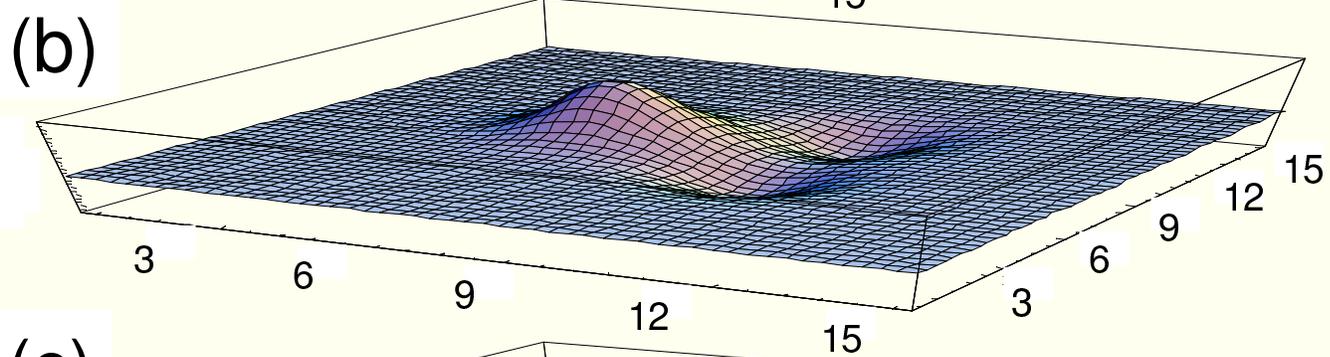
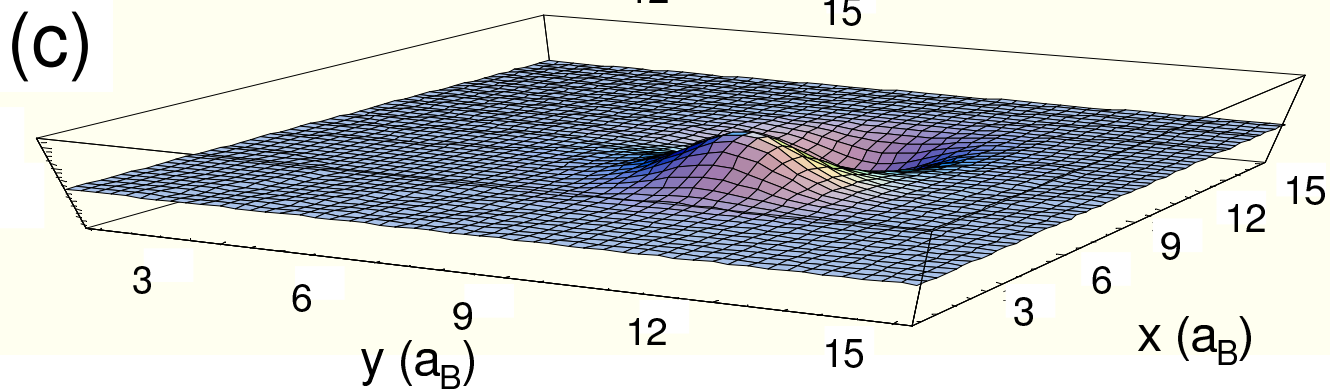
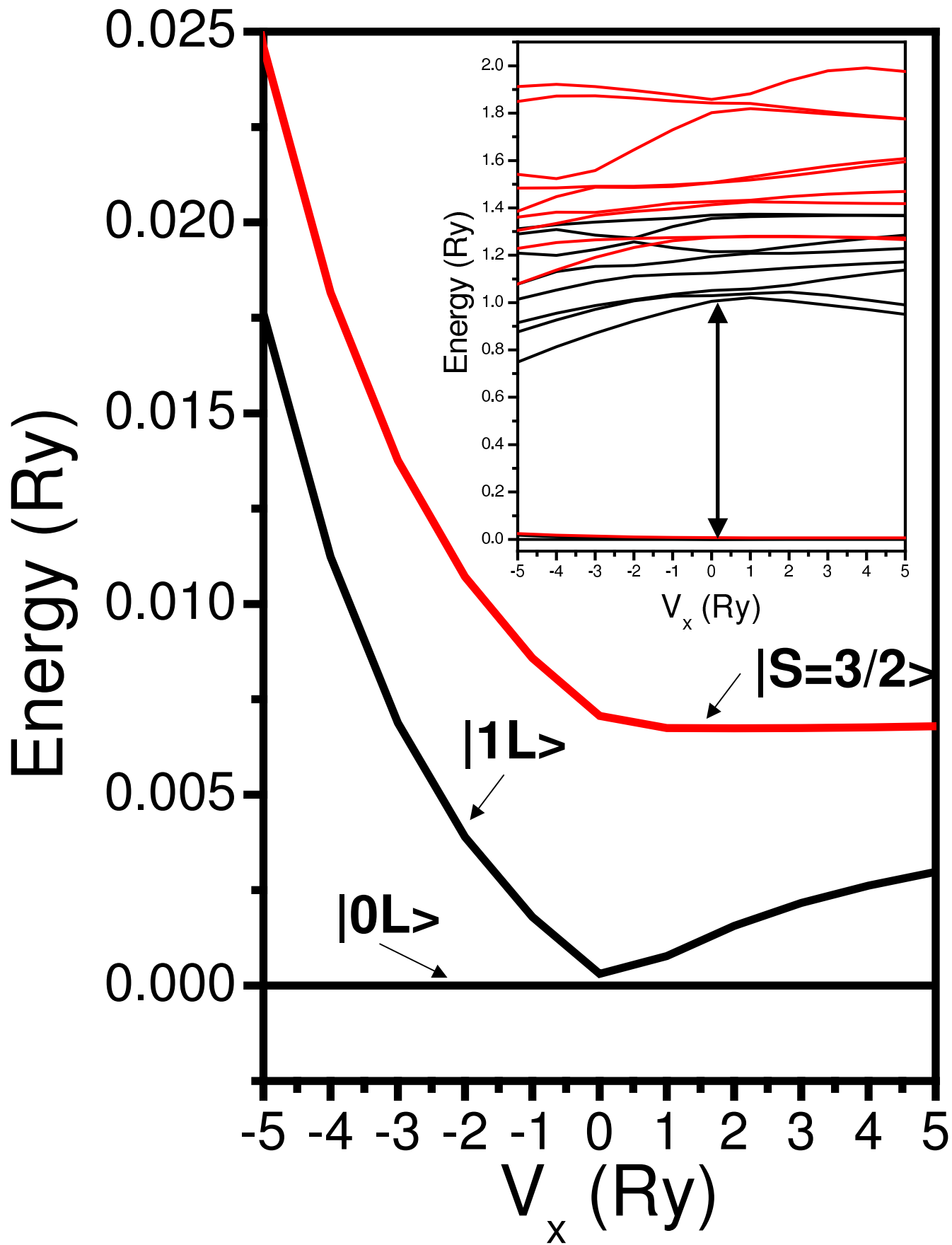
\end{document}